\documentclass[11pt]{elsarticle}
\usepackage{amsmath}
\usepackage{graphicx,setspace}
\usepackage{amsfonts,amsmath, amssymb, amsthm}
\usepackage{mathrsfs}
\usepackage{epstopdf}
\usepackage{float}
\usepackage{lineno,hyperref}
\usepackage{color}
\usepackage{subfigure}
\usepackage{bm}
\usepackage{subfigure}
\modulolinenumbers[5]
\numberwithin{equation}{section}

\journal{Applied Mathematics and Computation}

\begin{document}

\begin{frontmatter}

\title{Transitions between Metastable States in a Simplified Model for the Thermohaline Circulation under Random Fluctuations\tnoteref{mytitlenote}}
\tnotetext[mytitlenote]{This work was partly supported by the NSFC grant 11801192.}
\author[mymainaddress,mysecondaryaddress]{Daniel Tesfay}
\ead{dannytesfay@hust.edu.cn}

\author[mymainaddress]{Pingyuan Wei}
\ead{weipingyuan@hust.edu.cn}

\author[mymainaddress,mythirdaddress]{Yayun Zheng}
\ead{yayunzh55@hust.edu.cn}

\author[myfourthaddress]{Jinqiao Duan\corref{mycorrespondingauthor}}
\cortext[mycorrespondingauthor]{Corresponding author}
\ead{duan@iit.edu}

\author[myfifthaddress]{J\"{u}rgen Kurths}
\ead{kurths@pik-potsdam.de}

\address[mymainaddress]{School of Mathematics and Statistics \& Center for Mathematical Sciences,  Huazhong University of Science and Technology, Wuhan 430074, China}
\address[mysecondaryaddress]{Department of Mathematics, Mekelle University, P.O.Box 0231, Mekelle, Ethiopia}
\address[mythirdaddress]{Wuhan National Laboratory for Optoelectronics, Huazhong University of Science and Technology, Wuhan 430074, China}
\address[myfourthaddress]{Department of Applied Mathematics, Illinois Institute of Technology, Chicago, IL 60616, USA}
\address[myfifthaddress]{Research Domain on Transdisciplinary Concepts and Methods, Potsdam Institute for Climate Impact Research, P.O. Box 601203, 14412 Potsdam, Germany}

\begin{abstract}

In this work we study the impact of non-Gaussian $\alpha$-stable L\'evy motion on transitions between metastable equilibrium states (or attractors) in a stochastic Stommel two-box model for thermohaline circulation (THC). By maximizing  probability density of the solution process associated with a nonlocal Fokker-Planck equation, we compute maximal likely pathways and identify corresponding maximal likely stable equilibrium states. Our numerical results indicate weakened THC may be induced by perturbation with very small noise intensity in certain range of stability index. Moreover, larger noise intensity and larger stability index induce weakened THC within shorter bifurcation time.

\end{abstract}


\end{frontmatter}

\section{Introduction}

In the tremendous variety of nonlinear complex dynamical systems, among which the thermohaline circulation (THC) is one to mention, Gaussian as well as non-Gaussian noise play a pivotal role on setting up the dynamical behaviour of the system. This large-scale circulation in the ocean is decisive in the global climate change. The THC is driven by fluxes of heat and freshwater across the sea surface and posterior interior mixing of heat and salt \cite{Rahmstorf2003}. Modeling the THC is a daunting puzzle due to the lack of, for instance, universal equation of state which connects the density of water to temperature and salinity and the complicated form of the domain which is bounded by the edges of various continents. Taking the system-level approach, \emph{i.e.}, as a system the ocean is just a reservoir filled with salt water, where the circulation is driven by the density (heat and salinity) difference, eases the impasse. The oceanic conveyor belt (or THC) carries  warm surface water, around 17 Sv in volume and 0.5-1.5$ \times 10^{15}W$ in heat on annual average, from low latitude (or equatorial) to high latitude (or polar) regime where it cools down, gets denser and sinks. The cold  water at  deeper levels, 2-3 km in depth, back from the polar to the equator regime as deep water-currents and surface water becomes more salty at the equator as water is removed by evaporation due to the greater heat which make it heavier and less salty  thereupon it upwells.  To examine this circulation Stommel pioneered a simple deterministic two-box model \cite{Stommel1961}. This idea has been further extended and improved through both conceptual and numerical models \cite{Djikstra2005}-\cite{Rahmstorf1996}. Actually box models are taken as the simplest setting to investigate stochastically forced systems. Climate records reveal incidence of abrupt climatic changes that might have some link with transition between stable equilibrium states of the THC \cite{Selez2001}. Uncertain processes interrelated with random atmospheric fluctuations, challenges corresponding to unresolved scales and strange mechanisms worth consideration since they may lead to switching between the equilibrium states. It is customary to regard noisy fluctuations as Gaussian \cite{Cessi1996}-\cite{Yongkai2014}.
Velez-Belchi \emph{et al.} in \cite{Selez2001} verified stochastic resonance with additive noise induces transitions between the different stable states of the THC. A pathwise analysis of slowly varying not too large a perturbation on box model for THC has shown the system spends significant amount of time in metastable equilibrium \cite{Bergund2002}. Multiple stable states and possible critical transitions is discussed in \cite{Good2018}.

Noisy fluctuations, meanwhile, are demonstrated to be non-Gaussian in some complex systems such as spectral analysis of paleoclimatic data \cite{Ditlevsen1999}, metastability in climate systems \cite{Larissa2017}, and bursty transition in gene expression \cite{Niraj}.
The effect of Gaussian noise on THC has been well dealt with, while the influence of non-Gaussian stable L\'evy noise on the dynamical behaviour of this system is scant.
It is more expedient to model these random fluctuations which portray haphazard jumps, discontinuity, heavy tail distribution and bursting sample paths by a non-Gaussian ($\alpha$-stable) L\'evy motion. With respect to $\alpha$-stable L\'evy noise, Wang \emph{et al.} in \cite{Wang2012} determined low variability salinity difference level domain by numerical simulation of mean exit time.

In this paper, we focus on the influence of non-Gaussian $\alpha$-stable L\'evy noise on stochastic
box model for THC by investigating maximal likely attractors of the maximal likely pathways, \emph{i.e.}, evolution of the maximizer of probability density function as time increases in stochastic phase portraits. The subtlety of phase portraits for stochastic differential equations (SDE) can be supplemented by reconveying these geometrical entities in terms  maximal likely pathways \cite{Duan2015}, \cite{Cheng2016}.
We will consider a simple model for THC as a scalar SDE in the following form:
\begin{equation}\label{scalar}
dY_t = f(Y_t)dt + \sigma dL_t^{\alpha},     \qquad   Y(0) = Y_{0},
\end{equation}
where $f(Y_t)$ is a deterministic vector field, $\sigma$ is the noise intensity, $\alpha$ is stability index and \{$L_t$\}, $t\ge0$ is a scalar non-Gaussian L\'evy motion.

The paper is structured as follows. Section 2 presents details of a stochastic Stommel model. Section 3 reviews maximal likely stable equilibrium states or maximal likely attractors.
Section 4 discusses numerical results of computations on maximal likely pathways under different parameters, viz, stability index, noise intensity and (nondimensional) freshwater flux.

\section{Model and Method}
\setcounter{equation}{0}

\subsection{Model}

\begin{figure}[h!]
\centering
\includegraphics[width=3in]{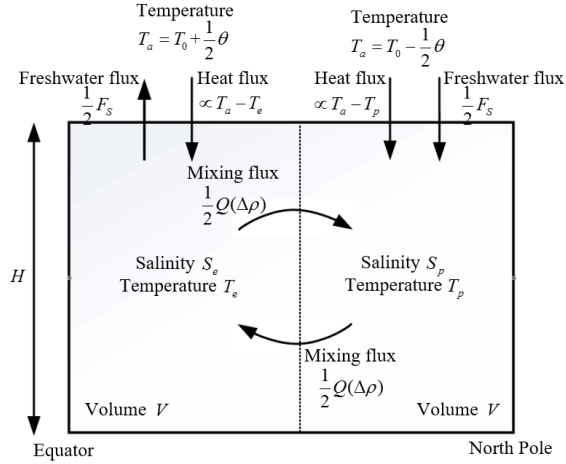}
\caption{ A variant of Stommel two-box model(\cite{Dijkstra2015}).}
\label{Stommel_model}
\end{figure}

An equatorial box and polar box with temperature $T_e$ and $T_p$, and salinity $S_e$ and $S_p$ respectively having the same volume $V$ are connected by advective flow and barter heat and freshwater with the atmosphere. Moreover, each container is individually forced at its boundary. The equatorial box bounces according to a time scale $t_r$ back to local atmospheric temperature $T_a$ = $T_0 + \frac{\theta}{2}$, where $\theta$ is the equator-to-pole atmospheric temperature difference and $T_0$ is a reference temperature. The polar box retains local atmospheric temperature, for the sake of symmetry, $T_a$ = $T_0 - \frac{\theta}{2}$. Freshwater is dispatched from equatorial into polar vessel by a laid down freshwater flux $\frac{F_s}{2}$ \cite{Dijkstra2013}. A sketch of a variant of Stommel's model is shown in Figure \ref{Stommel_model}.

Following \cite{Cessi1996}, we define the heat and salinity balances by the equations:

 \begin{equation}\label{four_equations}
 \begin{aligned}
  \dot{T_e} &= -{t_r}^{-1}(T_e-(T_0 + \frac{\theta}{2})) - {\frac{1}{2}}Q(\Delta\rho) (T_e-T_p), \\
  \dot{T_p} &= -{t_r}^{-1}(T_p-(T_0 - \frac{\theta}{2})) - {\frac{1}{2}}Q(\Delta\rho) (T_p-T_e),\\
  \dot{S_e} & = {\frac{F_S}{2H}}S_0-{\frac{1}{2}}Q(\Delta\rho) (S_e-S_p),  \\
  \dot{S_p }&= -{\frac{F_S}{2H}}S_0-{\frac{1}{2}}Q(\Delta\rho) (S_p-S_e).
 \end{aligned}
 \end{equation}

\begin{tabular} {llll}
\hline Parameter & Meaning & Value & Unit \\ \hline  $t_r$ & temperature relaxation time scale & 25  &  days  \\ H & mean ocean depth & 4,500  &  $m$  \\$t_d$ & diffusion time scale & 180  &  years \\$t_a$ & advective time scale & 29  &  years \\q & transport coefficient & $1.92  \times  10^{12}$  &  $m^3s^{-1}$\\  V & ocean volume & $ 300  \times 4.5 \times 8,250$  &  $km^3$\\  $\beta_T$ & thermal expansion coefficient & $10^{-4}$  &  $K^{-1}$  \\$\beta_S$ & haline contraction coefficient & $7.6 \times 10^{-4}$  &  $-$  \\  $S_0$ & reference salinity & 35  &  $g kg^{-1}$  \\ $\theta$ & meridional temperature difference & 25  &  $K$  \\ \hline \end{tabular}

Table 1. Parameters of the stochastic Stommel box model \cite{Dijkstra2013}\\

Density differences are essential for the circulation to take place. The density $\rho$ of a box, which is affected by $T$ and $S$ in an opposite way, is approximated by a linear equation of state
\begin{equation} \frac{\rho}{\rho_0} = 1 + \beta_S(S-S_0)-\beta_T(T-T_0),\nonumber \end{equation}
where $\beta_T$ and $\beta_S $ are (positive) constant thermal expansion and haline contraction coefficients, respectively. We may choose, for simplicity, a positive transport function which is independent of the direction of $\Delta\rho $. This is viable because in the closed advection of water between two vessels ($Q$), the quantity of equatorial water moving into the poles and vice versa keeps unaltered in both directions of circulation. Thus,\\
\begin{equation} Q(\Delta\rho)= \frac{1}{t_d} + {\frac{q}{V}} (\frac{\Delta\rho}{\rho_0})^2 ,\nonumber \end{equation}
where $t_d$ is diffusive mixing time scale between the two vessels that would occur when there is no density difference.
Subtracting the equations in (\ref{four_equations}) and defining $\Delta S \equiv  S_e - S_a $ and $\Delta T \equiv  T_e - T_a $ yields a coupled pair of equations for the time evolution of temperature and salinity differences between the vessels:

\begin{equation}\label{2eqns_dimensional}
 \begin{aligned}
  \frac{d\Delta T}{dt} &= -{t_r}^{-1} (\Delta T - \theta) - Q(\Delta\rho)\Delta T, \\
  \frac{d\Delta S}{dt} &= {\frac{F_S}{H}}S_0 - Q(\Delta\rho)\Delta S.
  \end{aligned}
 \end{equation}
With the scales $\Delta T \equiv x \theta, \Delta S \equiv \frac {y\beta_T\theta}{\beta_s}$ and time scaled with $t_d$,  we recast (\ref{2eqns_dimensional}) as the following coupled dimensionless system of equations.

\begin{equation}\label{2eqns_nondimensional}
 \begin{aligned}
  dx &= (-{\beta}(x-1) - x[1 + \mu^2(x - y)^2])dt, \\
  dy &= (F - y[1 + \mu^2(x - y)^2])dt.
  \end{aligned}
 \end{equation}
Where $\beta = \frac {t_d}{t_r}$ measures temperature restoring tensility, $\mu^2 = \frac{qt_d(\beta_T\theta)^2}{V}$ is strength of the buoyancy-driven convection between the two boxes relative to the diffusive mixing. The parameter $ F = {\frac{\beta_S S_0 t_d}{\beta_T \theta H}}F_S$ represents dimensionless freshwater forcing. The physical meaning of $\beta \gg 1 $ is that mixing between the boxes temperatures is sluggish compared to the rapid parity of each box's temperature with its local temperature forcing. Salinity leads to non-linearity which causes the existence of multiple equilibria and thresholds in the THC.\\

Due to the dominance of the $\beta$ term in equations (\ref{2eqns_nondimensional}) the temperature difference $x$ changes very little, and we can thus take $x$ as $ x = 1 + \mathcal{O}(\beta^{-1}) $. This leaves us with an ODE in $y(t)$,
\begin{equation}\label{ode}
dy = (F - y[1 + \mu^2(1 - y)^2])dt.
\end{equation}
Suppose that $F = \bar{F}$ is constant. The time evolution of $y$ can be represented by a potential function $ V(y)$ as $ dy = -V'(y)dt$ where $ V(y) = -{\bar{F}}y + \frac{1+\mu^2}{2}y^2 -{\frac{ 2\mu^2}{ 3}y^3 } + \frac{\mu^2}{4}y^4 $.  The strength of the transport between boxes is governed by density difference $ |\Delta \rho| = \rho_0 |\beta_S\Delta S - \beta_T\Delta T|$ which in dimensionless form is $|y - x|$. In certain parameter ranges $ V(y)$ has two stable states: one with larger northward heat transport, usually referred to as the \emph{on}-state, where the salinity difference $ y $ is small, and other with weak (or even reversed) circulation, where the difference $ y $ is large, called the \emph{off}-state, depending on differences in the initial conditions only.
After Cessi's valuation of parameter values \cite{Cessi1996}, \cite{Dijkstra2013}  $\bar{F} = 1.1$, $\mu^2 = 6.2$, $ V(y)$ is a double-welled potential with two stable minima at $y_{-} = 0.24$ and $y_{+} = 1.07$ and an unstable maximum at $y_{u}= 0.69$ as can be seen in Figure \ref{Stommel_model}.\\

\begin{figure}[h!]
\centering
\includegraphics[width=3in]{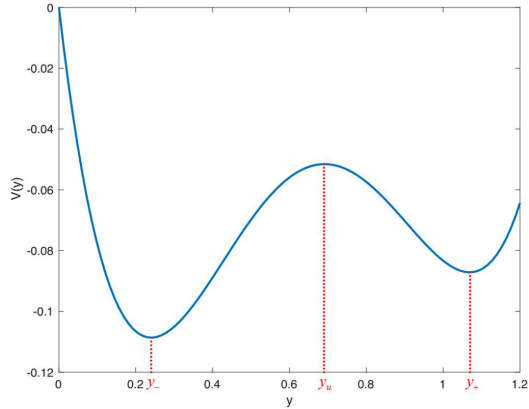}\\
\caption{ The potential $V$ as a function of the salinity difference $y$.}
\label{Potential}
\end{figure}

Numerical simulations of the THC disclose its deep transport, stability and variability delicacy , especially with respect to freshwater flux \cite{Gregory},\cite{Timmermann}. A multiplicity of models has displayed that the competition between thermal and saline forcing results in multiple equilibria. Clearly, the deterministic system has two stable steady states and the random driving moves the temperature and the salinity away from initial equilibrium. So a perturbation is certainly needed to induce a transition between both states. These perturbations are provided by background noise such as the freshwater flux. The fluctuations in the freshwater flux or in the external salinity inputs lead $F$ to vary following the difference between salinity increment rate at the equator and corresponding decline at north pole. Further, $F$ may be parameterized as a sum of time independent mean component $\bar{F}$ and a noisy fluctuating process $ \frac {dL_t}{dt}$. As a result, we come out with the SDE:
\begin{equation}\label{sde}
dY_t = (\bar{F}-Y_t[1 + \mu^2(1-Y_t)^2])dt + \sigma dL_t.
\end{equation}

\subsection{$\alpha$-stable L\'evy process}

A stable distribution $S_{\alpha}(\mu,\beta,\gamma)$ is the distribution for a stable random variable \cite{Duan2015},  where the parameters $\alpha \in (0,2),\mu \in [0,\infty),\beta \in [-1,1]$ and $\gamma \in (-\infty, \infty)$ respectively are indices of stability, scale, skewness and shift.

An $\alpha$-stable scalar L\'evy motion $L_t^{\alpha}, \alpha \in (0,2)$ is a non-Gaussian stochastic process satisfying the conditions:\cite{Duan2015}, \cite{Applebaum2009}-\cite{Sato1999}\\
i) $L_0^{\alpha} = 0$, almost surely;\\
ii) Independent increments: For each $n \in \mathbb{N}$ and each  $ 0 \le t_1 < t_2 < \cdots  < t_{n-1} < t_n < \infty $, the random variables $ L_{t_2}^{\alpha} - L_{t_1}^{\alpha} , \cdots  ,L_{t_n}^{\alpha} - L_{t_{n-1}}^{\alpha}$ are independent;\\
iii) Stationary increments: $ L_t^{\alpha} - L_s^{\alpha} $ and $ L_{t-s}^{\alpha} $ have the same distribution $S_{\alpha}((t-s)^{\frac{1}{\alpha}},0,0)$;\\
iv) Stochastically continuous sample paths:  Sample paths are continuous in probability. For all $\delta > 0 $, all $ s \ge 0; \mathbb {P}(|L_t^{\alpha} - L_s^{\alpha}|) > \delta)\to 0$ as $ t \to s$. \\ \\
The jump measure of $L_t^{\alpha}$ is
\begin{equation}\nu_\alpha (dz) = C_\alpha |z|^{-(1+\alpha)}dz, \nonumber \end{equation} where $ C_\alpha = \frac{\alpha}{2^{1-\alpha}{\sqrt\pi}}\frac{\Gamma(\frac{1+\alpha}{2})}{\Gamma(1-\frac{\alpha}{2})} $ is stability constant.

Brownian motion, $B_t$, is a symmetric 2-stable process. $B_t$ has independent and stationary increments, continuous sample paths almost surely and has normal distribution $\mathcal{N}(0,t)$.

\subsection{Method}

We will consider a simple model for THC as a scalar SDE with additive noise, \emph{i.e.} the intensity of the noise is independent of the state of the THC, in the form
\begin{equation}\label{additive_noise}
dY_t = f(Y_t)dt + \sigma dL_t^{\alpha},     \qquad   Y(0) = y_0,
\end{equation}
where $f(Y_t) = \bar{F}-Y_t[1 + \mu^2(1-Y_t)^2]$ is a deterministic vector field, $\sigma$ is the noise intensity and \{$L_t^{\alpha}$\}, $t\geq0$, is a symmetric scalar $\alpha$-stable non-Gaussian L\'evy motion.\\
The generator A of the solution process $Y_t$ for the SDE (\ref{additive_noise}) with triplet $(0,0,\nu_\alpha)$ is
\begin{align} A \varphi (y,t) = f(y)\varphi'(y,t)+ \int_{\mathbb {R}^1\backslash\{ 0 \}}[\varphi (y + \sigma z,t)-\varphi (y,t)]\nu_{\alpha}(dz)\nonumber\\
                     = f(y)\varphi'(y,t)+\sigma^\alpha\int_{\mathbb {R}^1\backslash\{ 0 \}}[\varphi(y + z,t)-\varphi(y,t)]\nu_{\alpha}(dz),&\end{align}
where $\varphi (y)\in C(\mathbb {R}^1)$ and the Fokker-Planck equation of (\ref{additive_noise}) in terms of the probability density function $p(y,t)$ for the solution process $Y_t$ given initial condition $Y_0 = y_0$ is \cite{Duan2015}:
\begin{equation}\label{fpe}
p_t(y,t) = A^*p(y,t),  \quad p(y,0)= \delta (y-y_0),
\end{equation}
here, $\delta$ represents the dirac function and $A^*$, the adjoint operator of $A$ in the Hilbert space $L^2(\mathbb {R}^1)$, can be computed by solving
\begin{equation}
 \int_{\mathbb {R}^1\backslash\{ 0 \}} A\varphi(y)\upsilon(y)dy = \int_{\mathbb {R}^1\backslash\{ 0 \}} \varphi(y)A^*\upsilon(y)dy, \nonumber
\end{equation}
for $\varphi, \upsilon $ in the domains of definition for the operators $A$ and $A^*$, respectively,
\begin{equation}
 A^*\upsilon(y)= \sigma^\alpha \int_{\mathbb {R}^1\backslash\{ 0 \}}[\varphi(y+z)-\varphi(y)]\nu_{\alpha}(dz). \nonumber
\end{equation}
Consequently, we obtain the following nonlocal Fokker-Planck equation
\begin{equation}\label{nonlocal_fpe}
 p_t = -(f(y)p(y,t))_y + \sigma^\alpha \int_{\mathbb {R}^1\backslash\{ 0 \}}[\varphi(y+z)-\varphi(y)]\nu_{\alpha}(dz).
\end{equation}
The SDE
\begin{equation}\label{Brownian}
 dY_t = f(Y_t)dt + \sigma dB_t,     \qquad   Y(0) = y_0
\end{equation}
is brought in if Brownian motion, which has a Fokker-Planck equation having the form of local partial differential equation (\ref{local_pde}), substitutes the L\'evy motion.
\begin{equation}\label{local_pde}
 p_t = -(f(y)p(y,t))_y + \frac{1}{2}\sigma^2p(y,t)_{yy},    \qquad p(y,0) = \delta(y-y_0).
\end{equation}
Numerical finite difference method as in \cite{Gao2016} and standard difference method are used in simulating equations (\ref{nonlocal_fpe}) and (\ref{local_pde}) respectively.

\section{Maximal Likely  Trajectories}
\setcounter{equation}{0}
Phase portraits provide geometric pictures for lower-dimensional deterministic dynamical systems. At a given time instant $t$, the maximizer of $y_m(t)$ of the probability density function $p(y,t)$ yields the most probable location of this trajectory at time $t$. The deterministic entity $y_m(t)$ follows the uppermost crest of the surface in the $(y,t, p)$-space leaving behind trace out of the so called the maximal likely trajectory starting at $y_0$,  as time goes on.

A state that attracts all nearby states is called a maximal likely stable equilibrium state. Maximal likely stable equilibrium states are dependent on non-Gaussianity index $\alpha$, noise intensity $\sigma$ and freshwater flux $\bar{F}$. The most probable phase portrait \cite{Duan2015}, \cite{Cheng2016}, an extension of phase portraits to envision stochastic dynamics, is the state space with representative maximal likely pathway including stable equilibrium states. Most probable phase portraits like the phase portraits are geometric entities. They depict sample pathways and stable equilibrium states that are maximal likely.

We examine the number and location of maximal likely stable equilibrium salinity difference states as a parameter varies. We may define bifurcation time as the time between the change in number of maximal likely equilibrium states. It is a time scale for the birth of a new most probable stable equilibrium state.

\section{ Result and Discussion }

In this section, we will discuss the number and value of maximal likely stable equilibrium states for the stochastic THC model (\ref{additive_noise}). For simplicity, we consider four maximum likely trajectories with initial points $y_0 = 0.2,0.4,0.8,1.5$, to both sides of the equilibria in the deterministic dynamical system. Due to the lengthy computation process time is capped to $ T =50 $.

Next we examine maximal likely trajectories when system parameters change.

\medskip
\textbf{As L\'evy noise parameters vary}:

The deterministic dynamical system is bistable in some range of freshwater flux parameter that contains $\bar {F} = 1.1$. A perturbation with very small noise intensity
on the deterministic system invokes an interesting phenomenon as the L\'evy noise index varies.\\
For $\sigma = 0.001$, in line with the existence of a transient state the interval of alpha stability can be partitioned in two.\\
i) $0.1 \lessapprox \alpha < 0.7$: In this interval of stability index, we have low salinity difference transient maximal likely stable equilibrium state. The transient state terminates since maximal likely salinity difference trajectories in the stochastic THC model merge by jumping to maximal likely metastable equilibrium state as time counts on. Bifurcation time is a time interval for the emergence of a new maximal likely stable equilibrium state. This time scale predicts transition from low salinity difference maximal likely state (or strong THC) to high salinity difference maximal likely state (or weak THC). Jump from low salinity
\begin{figure}[!htb]
\subfigure[]{ \label{fig3a}
\includegraphics[width=0.5\textwidth]{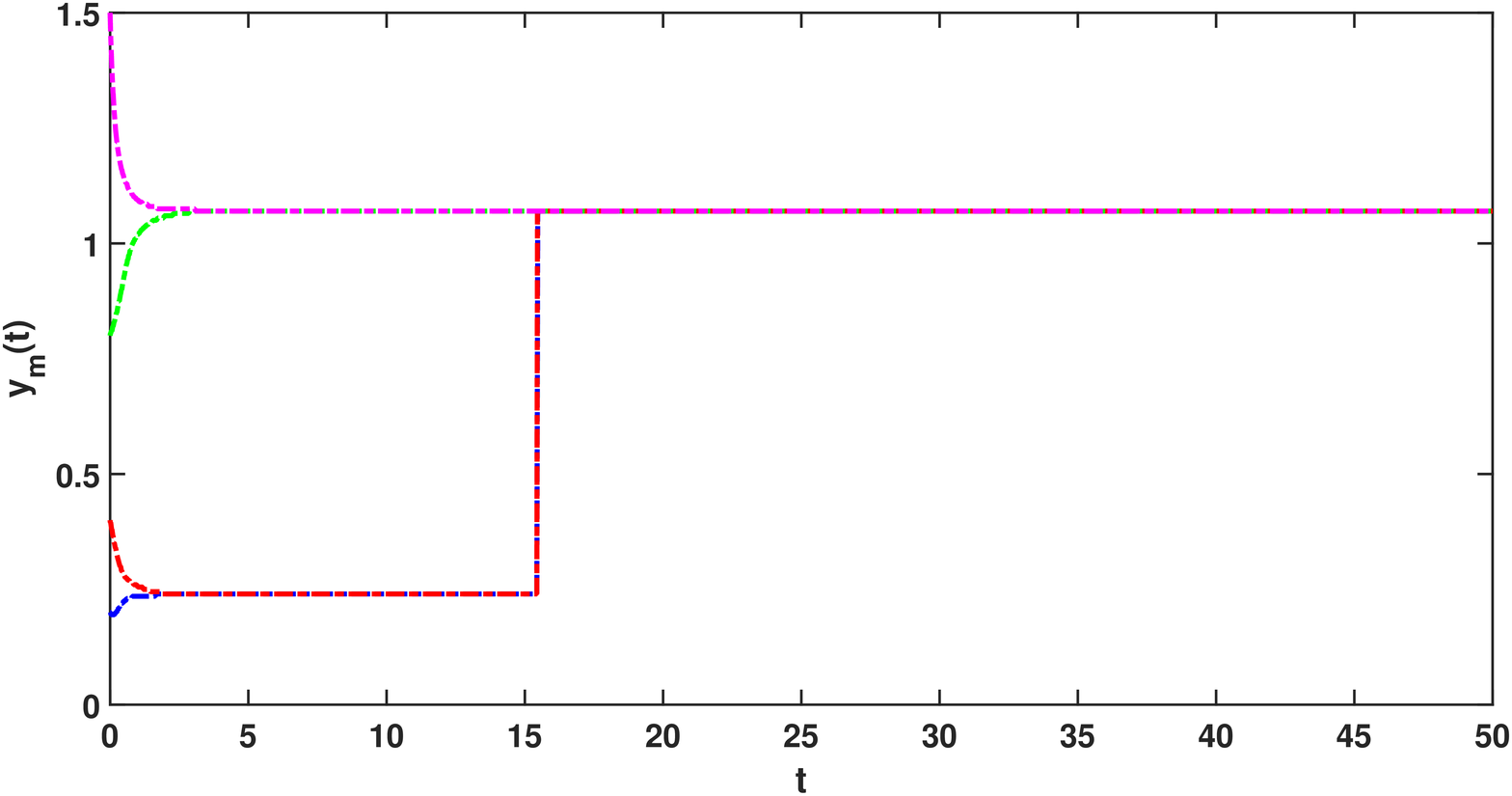}}
\subfigure[]{ \label{fig3b}
\includegraphics[width=0.5\textwidth]{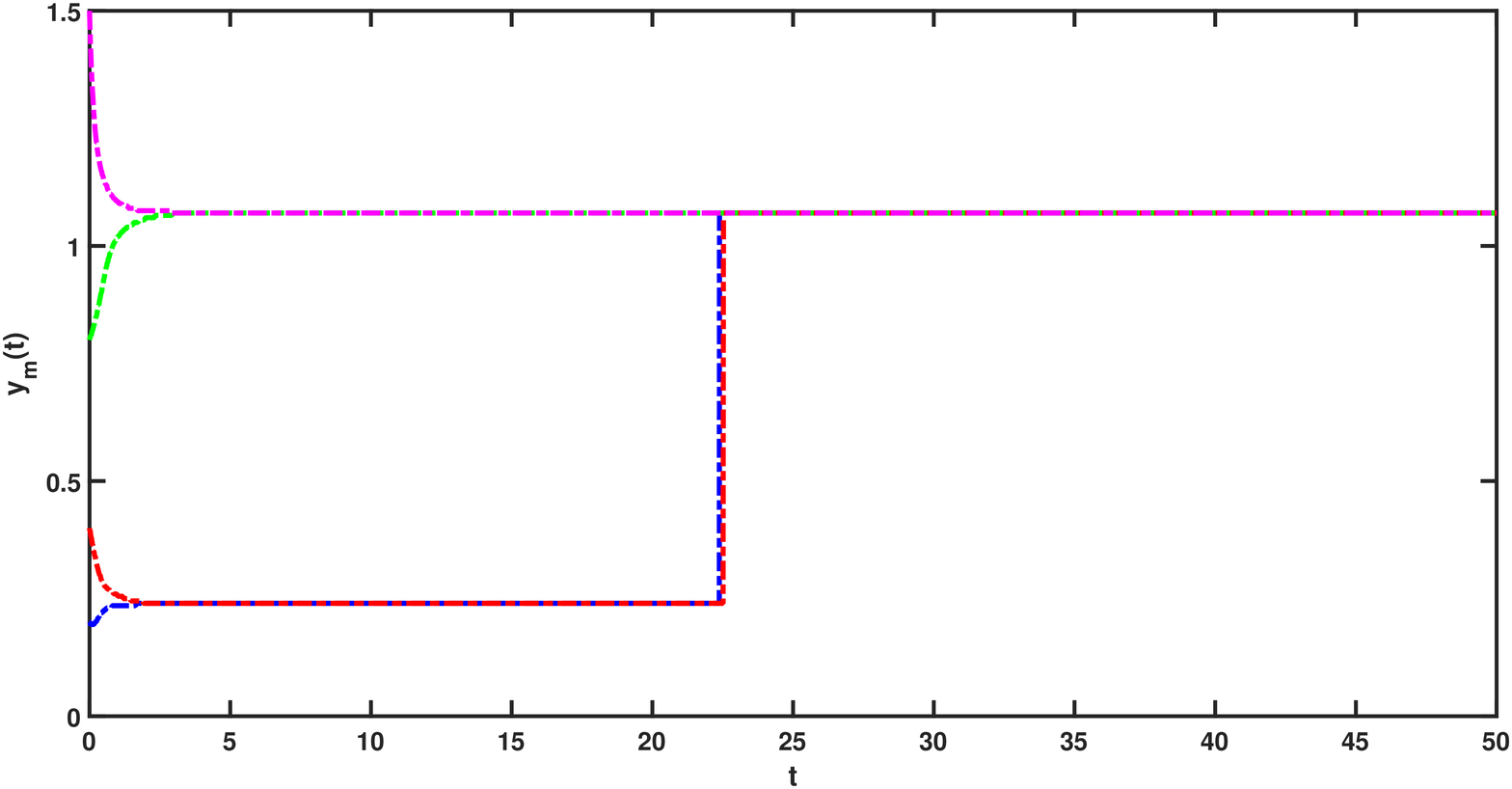}}
\subfigure[]{ \label{fig3c}
\includegraphics[width=0.5\textwidth]{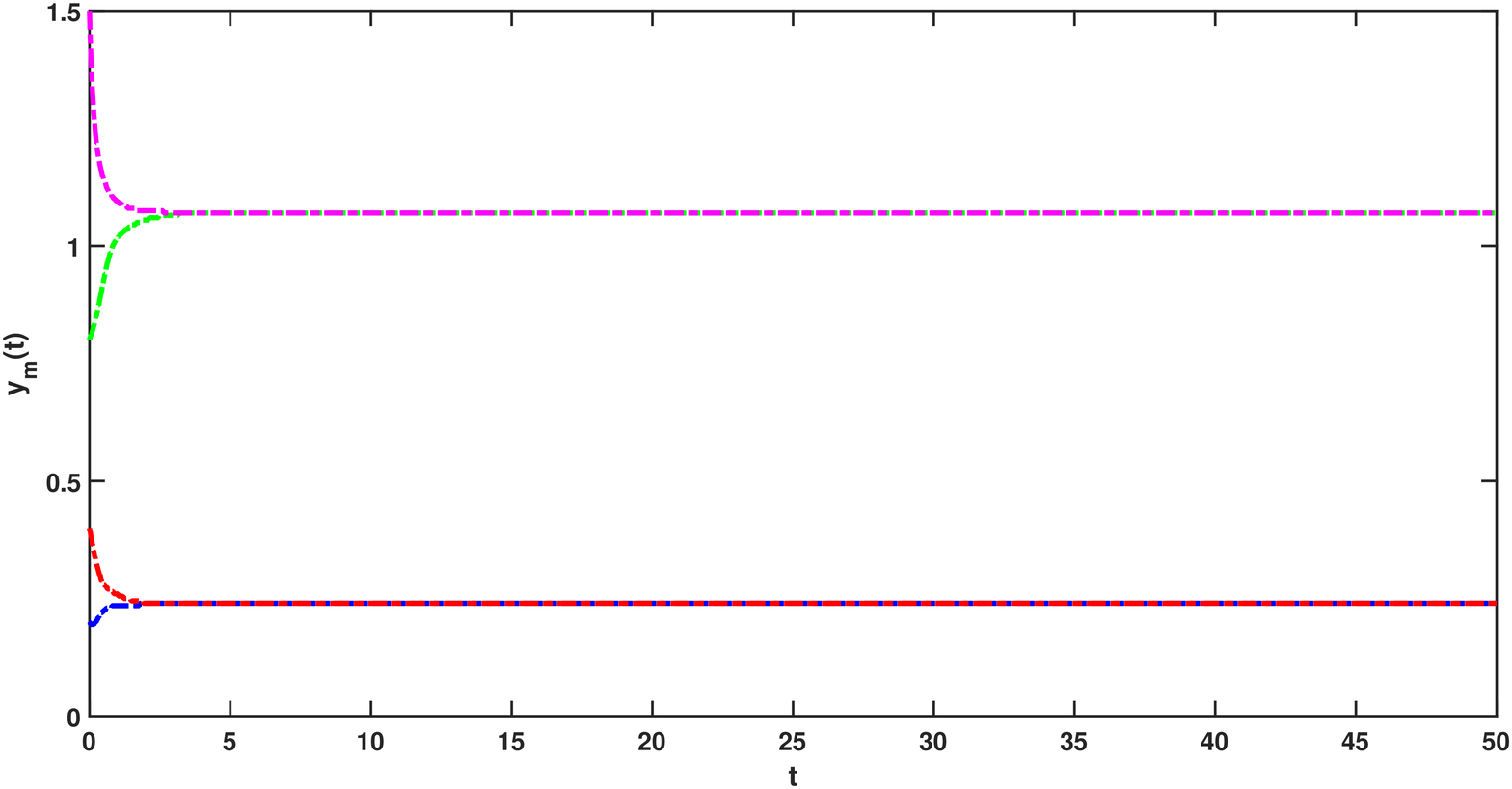}}
\subfigure[]{ \label{fig3d}
\includegraphics[width=0.5\textwidth]{F1p1e0p001a0p1t50.eps}}
\caption{(Color online) Maximal likely stable equilibrium salinity difference states 0.24 and 1.07 for $\bar{F}= 1.1, \sigma = 0.001$ and  (a) $\alpha = 0.15$. (b) $\alpha = 0.3 $. (c) $\alpha = 1$. (d) $\alpha = 2$. }
 \label{Figure_3}
\end{figure}
difference
to high salinity difference maximal likely equilibrium state comes out at different bifurcation time as index of stability in the interval varies. As can be seen from figure \ref{fig3a}, bifurcation time attains its minimum at $ t \thickapprox 15.4683$  when $\alpha\thickapprox 0.15$. This duration should be multiplied by the diffusion time, taken here to be $t_d = 180 $ years as in Table (1). Corresponding dimensional value of this minimum time is therefore about 2,784 years. As shown in figure \ref{fig4a}, bifurcation time decreases, but increases after it attains minimum  as $\alpha$ increases in this interval. In figure \ref{fig3b} bifurcation time is at $t \thickapprox 22.5279$ (about 4,055 years) when $\alpha = 0.3$. The transient maximum likely equilibrium state lasts for relatively short time interval before it ends jumping up to the maximal likely metastable equilibrium state. Next, we turn our focus on discussing why this jump takes place. An $\alpha$-stable process moves predominantly by big jumps as $\alpha$ nears to 0, and small jumps override as $\alpha$ is approaches 2. The jumps that occur in the maximal likely trajectories due to the L\'evy noise perturbation in this interval are seldom but big in size, as a result
the salinity difference trajectories originating from a small neighbourhood of the deep well have a substantial probability of bypassing the barrier wall to land in the shallow well. It can also be clearly seen that the bifurcation time increases or equivalently transient maximal likely stable equilibrium state lasts longer, with the increase in $\alpha$ bound to the interval under consideration.
Since $y$ is dimensionless salinity difference, it is proportional to the salinity difference between the equatorial and polar regions, higher $y$ values imply reduced density difference.

Thus, higher values of $ y $  may consequently lead to diminished THC. This implies noise with such a small noise intensity and some range of values of stability indices may invoke weakened THC.\\
For the purpose of comparison, the influence of Gaussian noise with the same noise intensity is shown in figure \ref{fig3d}. The stochastic system (\ref{Brownian}) is bistable. The reason for transition in the L\'evy noise case may be attributed to the jump in the evolution of salinity difference trajectories.

\begin{figure}[!htb]

\subfigure[]{ \label{fig4a}
\includegraphics[width=0.45\textwidth]{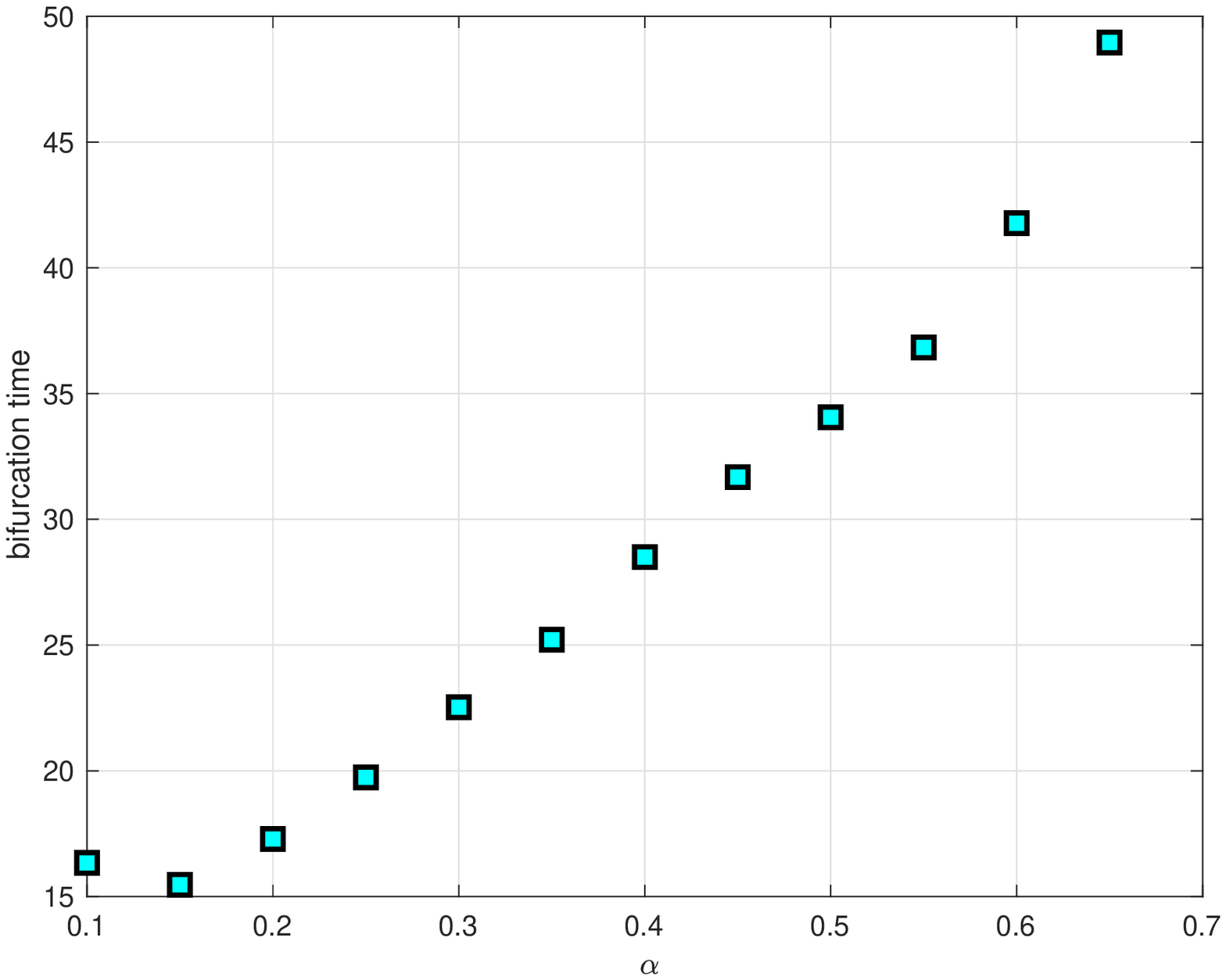}}
\subfigure[]{ \label{fig4b}
\includegraphics[width=0.45\textwidth]{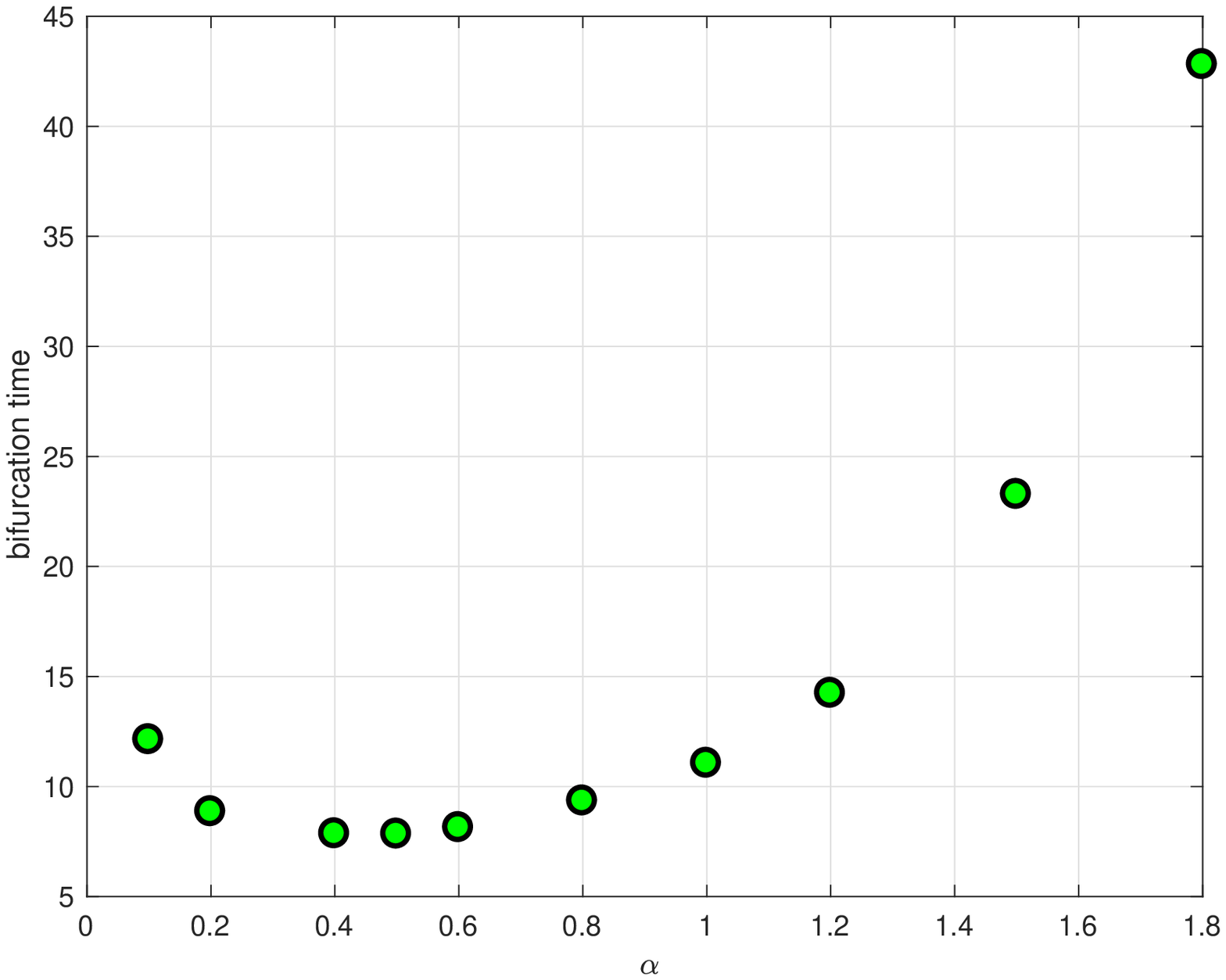}}

\caption{(Color online) The stochastic model experiences jumps from \emph{on} to \emph{off} state when $\bar{F} = 1.1$. (a) The system is monostable when $\alpha < 0.7$, and the minimum bifurcation time is $ t \thickapprox 15.4683$ when $\alpha = 0.15$ and $\sigma = 0.001$. (b) For $\sigma = 0.1$ bifurcation time is $ t \thickapprox 7.8275$ when $\alpha = 0.5. $}
\label{Fig4}
\end{figure}

ii) $0.7 \lessapprox \alpha < 2$: Our numerical results display the stochastic Stommel two-box model for THC in the SDE (\ref{additive_noise}) has two maximal likely stable equilibrium states.
The stable equilibrium states in the absence of noise match with these maximal likely stable states after perturbation. The stochastic system in this interval of stability index is, therefore, not affected by the noise. In figure \ref{fig3c}, for instance, the two most likely stable salinity difference are at 0.24 and 1.07.
An implication of this phenomenon in the perturbed THC model is the possibility that there is neither a shift in potential wells from the original location nor a state transition. Intensity of the noise is not large enough to stir a slight change in the maximal likely equilibrium states and the jumps occurring in the maximal likely trajectories are lower than the height of the barrier wall between the deep potential well and the shallow potential well (see figure \ref{Potential}). \\
Numerical experiments reveal the range of stability index where maximum likely transient equilibrium state prevails gets wider as noise intensity increases from  $0.001$ to $0.1$ as shown in \ref{Fig4}. In figure \ref{fig4a}, the system is monostable  when stability index is below 0.7 whereas the system is bistable when stability index exceeds 1.9 in figure \ref{fig4b}. Mean while, the bifurcation time is affected adversely by this noise intensity increment. The minimum bifurcation time decreases from 15.4683 (about 2,784 years) in figure \ref{fig4a} to 7.8275 (about 1,409 years) in figure \ref{fig4b}. Notice that the stability index as well increases from 0.15 to 0.5. This result hints that the transition from weak to strong THC may be enhanced as the L\'{e}vy noise parameters increase.

\textbf{As value of  $ \bar{F} $  varies}:

$ \bar{F} $ is of particular interest as it monitors the strength of the freshwater forcing and determines the stable states and transitions between stable states of THC. In the absence of noise, the model is bistable in the interval  $ 0.96 \lessapprox \bar{F} \lessapprox 1.3$, otherwise there is only one stable state: \emph{on}-state for $\bar{F}\lessapprox 0.96$ or \emph{off}-state when $\bar{F}\gtrapprox 1.3$.
It is desirable to examine how this stochastic THC model is influenced as freshwater flux parameter varies.
We consider very small values of the parameters $\alpha = 0.05$ and $\sigma = 0.01$ and the same initial conditions.
When the value of the freshwater flux parameter varies while the system is under the influence of non-Gaussian L\'evy noise, our numerical results show that, for $ \bar{F} = 1.1 $  there are two maximal likely stable equilibrium states at 0.24 and 1.07.

\begin{figure}[!htb]
\subfigure[]{ \label{fig5a}
\includegraphics[width=0.5\textwidth]{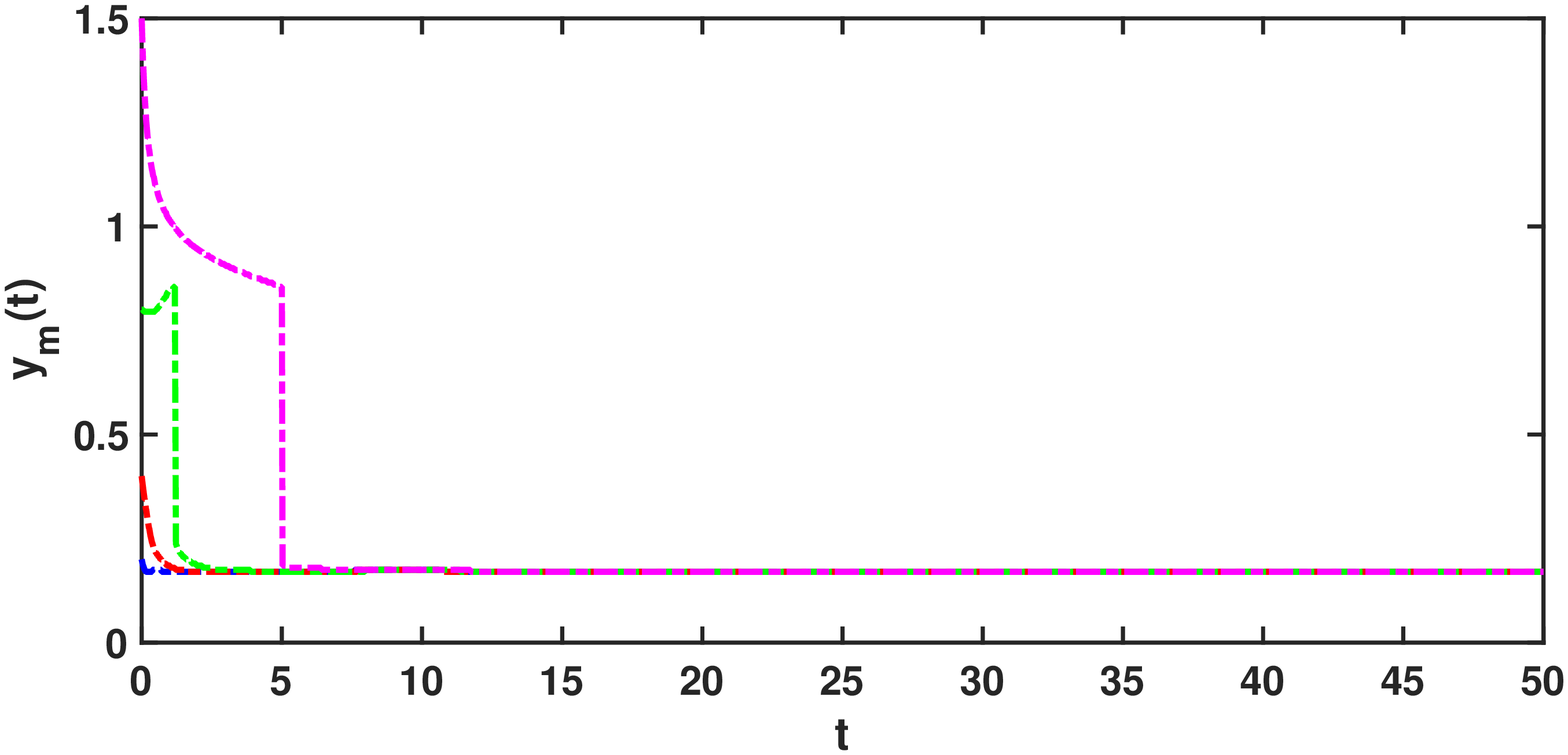}}
\subfigure[]{ \label{fig5b}
\includegraphics[width=0.5\textwidth]{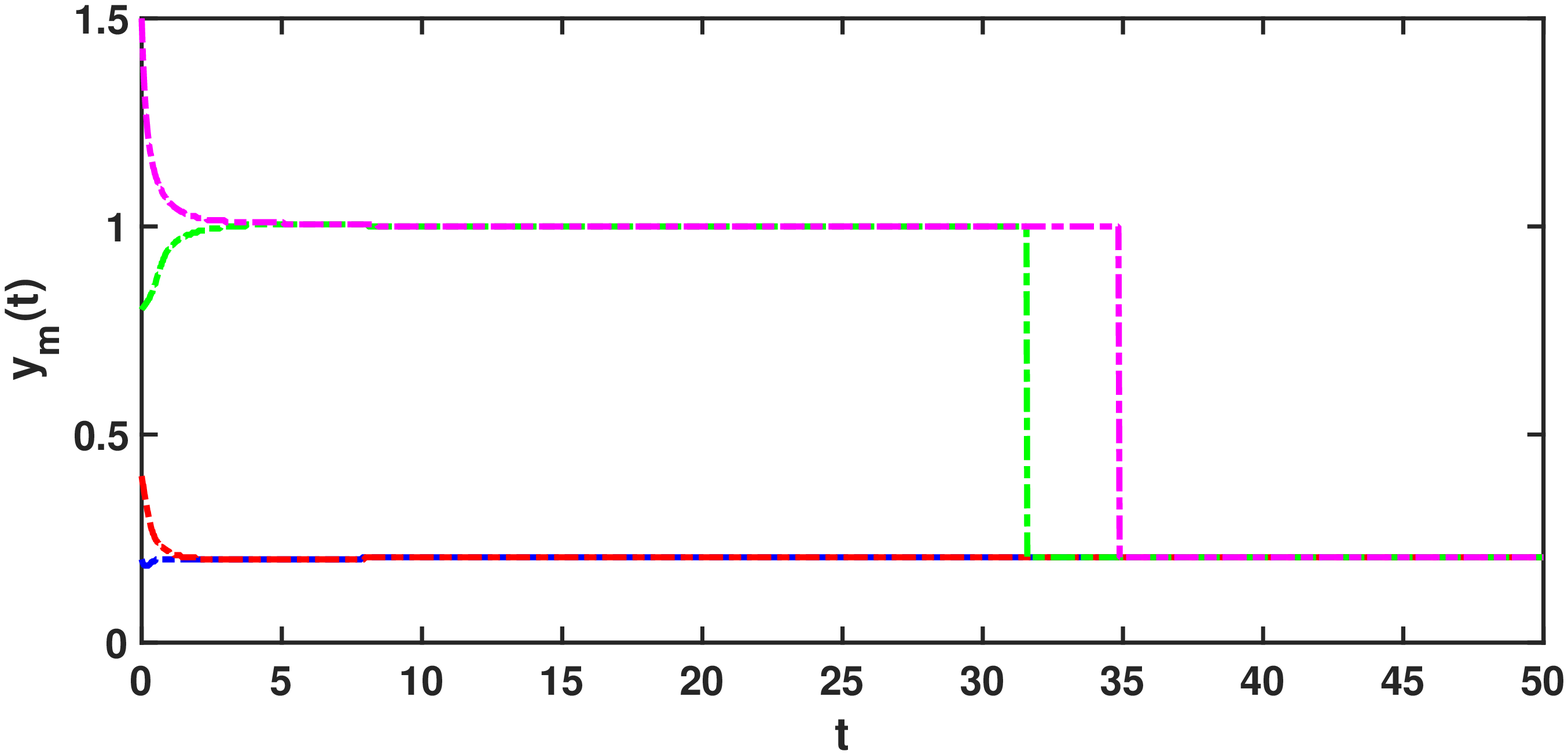}}
\subfigure[]{ \label{fig5c}
\includegraphics[width=0.5\textwidth]{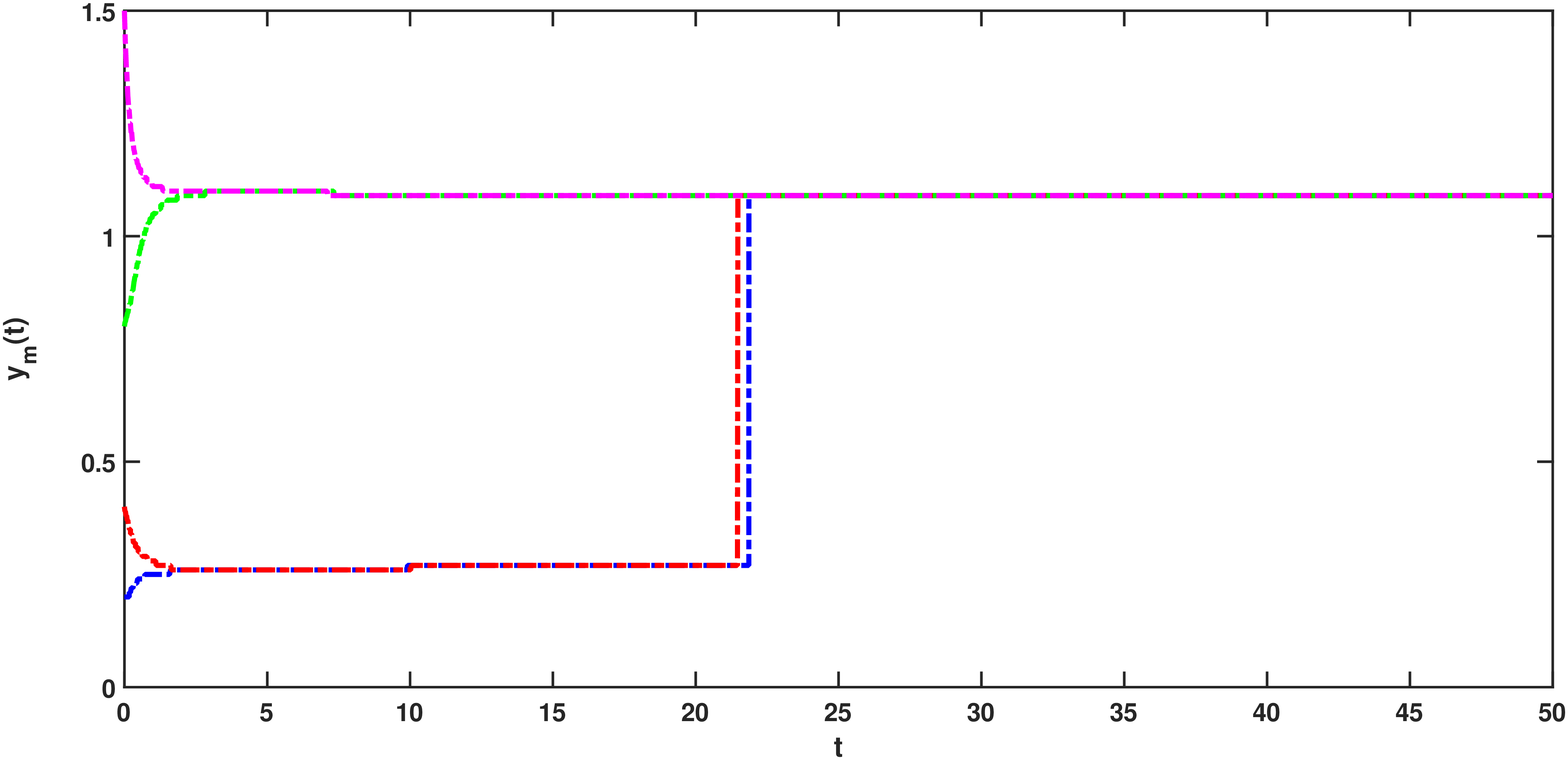}}
\subfigure[]{ \label{fig5d}
\includegraphics[width=0.5\textwidth]{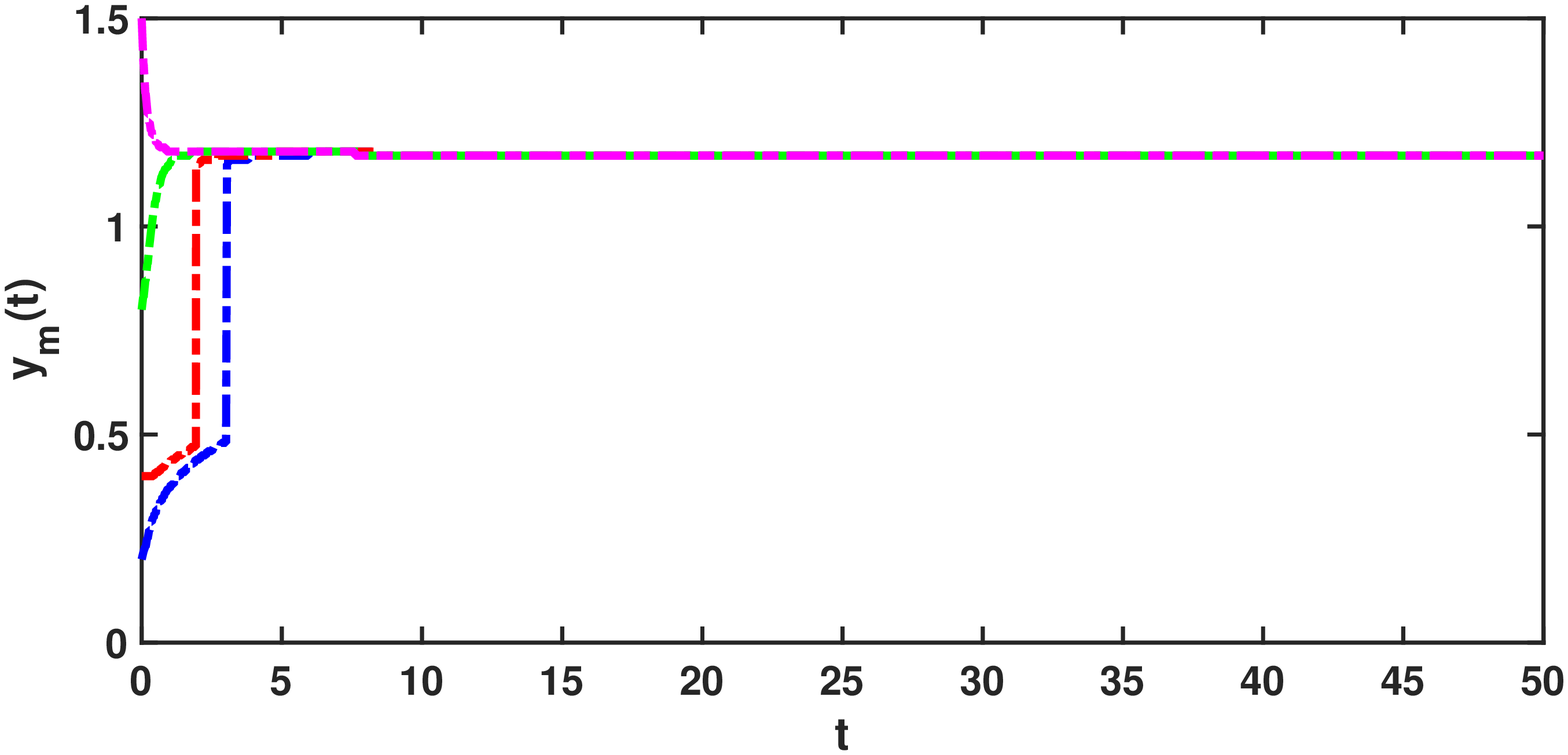}}

\caption{(Color online) Maximal likely stable equilibrium salinity difference states for  $\sigma = 0.01$, $\alpha = 0.05$ and (a) $\bar{F} = 0.9$. (b) $\bar{F} = 1$. (c) $\bar{F} = 1.15$. (d) $\bar{F} = 1.4.$}
 \label{Fig_8}
\end{figure}

It is worth noting that maximal likely stable equilibrium states of the stochastic system under the given freshwater flux parameter, stability constant and noise intensity remain the same as that of the deterministic system.
When $ \bar{F} = 0.9$, as in figure \ref{fig5a}, the stochastic system has only one maximal likely stable equilibrium state at 0.17. Similarly, at 0.185 for $ \bar{F} = 0.95 $. Actually in these, the stochastic system is monostable and these maximal likely low salinity difference states correspond to strong THC.
Existence of transient maximal likely equilibrium state is an interesting event that takes place when $ \bar{F} = 1 $, and $ \bar{F} = 1.15 $. In figure \ref{fig5b}, the transient maximum likely high salinity difference equilibrium state 1 merges to the maximal likely low salinity difference stable equilibrium state 0.205 at the bifurcation time $t = 34.9$ (about 6,282 years) when $ \bar{F} = 1 $. For $\bar{F} = 1.15$, the transient maximal likely equilibrium state is at 0.27 and the maximal likely metastable equilibrium state is at 1.09. The jump is from high to low salinity difference state and it takes place at bifurcation time $ t = 21.8606$ (about 3,934 years) as in figure \ref{fig5c}. When there is a transient state, numerical simulation results show bifurcation time increases and the transient maximal likely stable equilibrium state is low salinity difference state or strong THC which finally merges to high salinity difference state or weak THC when $ \bar{F} < 1.1 $, while if $ \bar{F} > 1.1$, bifurcation time decreases and the transition is reversed. The system has maximum likely high salinity difference equilibrium state (or weak THC) 1.17 if $\bar{F} = 1.4$ as shown in \ref{fig5d}. \\
Abrupt climate changes may have some degree of relationship with
sudden fluctuations of the THC due to vast freshwater flux caused by break up and melting of iceberg and glaciers. This conveyor belt transfers huge amount of heat from low to high latitudes regulating the global climate by redistributing heat across the planet.
Weak THC may have far reaching consequences in water sources, precipitation for farming and well-being of low latitude ecosystems. On the other hand, it may lead to cooler ocean temperature in certain pockets of high latitudes and warmer temperature in other pockets with potential to extra sea level rise along coasts and ascends sea temperature adversely affecting the oceanic ecology.

\section*{Acknowledgements}

 The authors would like to thank Xiujun Chen, Hui Wang, Xiaoli Chen and Ziying He for helpful discussions. This work was partially supported by NSFC grant 11801192.


\section*{References}

 
\end{document}